\documentclass[journal,12pt,onecolumn,draftclsnofoot,]{IEEEtran}
\IEEEoverridecommandlockouts
\usepackage{amsmath,amssymb,amsfonts}
\usepackage{algorithmic}
\usepackage{algorithm}
\usepackage{array}
\usepackage[caption=false,font=normalsize,labelfont=sf,textfont=sf]{subfig}
\usepackage{textcomp}
\usepackage{stfloats}
\usepackage{url}
\usepackage{verbatim}
\usepackage{graphicx}
\usepackage{ntheorem}
\usepackage[algo2e]{algorithm2e} 
\usepackage{cite}
\usepackage{lipsum}
\usepackage{mathtools}
\usepackage{cuted}
\usepackage{setspace}
\usepackage{multicol}
\usepackage{epsfig}
\usepackage{epstopdf}
\usepackage{textcomp}
\usepackage{xcolor}
\usepackage{booktabs, multirow}
\usepackage{siunitx}

\usepackage{kantlipsum}

\usepackage{float}

\newcommand{\argmax}{\operatornamewithlimits{argmax}}

\makeatletter

\newcommand{\Rmnum}[1]{\expandafter\@slowromancap\romannumeral #1@}
\def\BibTeX{{\rm B\kern-.05em{\sc i\kern-.025em b}\kern-.08em
    T\kern-.1667em\lower.7ex\hbox{E}\kern-.125emX}}

\begin{document}

\title{Distributed Pilot Assignment for Distributed Massive-MIMO Networks}

\author{\IEEEauthorblockN{Mohd Saif Ali Khan*, Samar Agnihotri* and Karthik R.M.$^+$}\\
\IEEEauthorblockA{*School of Computing \& EE, Indian Institute of Technology Mandi, HP, India}\\
\IEEEauthorblockA{$^+$Ericsson India Global Services Pvt. Ltd., Chennai, TN, India}\\
Email: saifalikhan00100@gmail.com, samar.agnihotri@gmail.com, r.m.karthik@gmail.com
}

%




\maketitle

\begin{abstract}
Pilot contamination is a critical issue in distributed massive MIMO networks, where the reuse of pilot sequences due to limited availability of orthogonal pilots for channel estimation leads to performance degradation. In this work, we propose a novel distributed pilot assignment scheme to effectively mitigate the impact of pilot contamination. Our proposed scheme not only has low convergence time, but it also enhances overall system performance with low signaling overhead. Results of extensive numerical simulations establish that the proposed scheme outperforms existing centralized and distributed schemes in terms of mitigating pilot contamination and significantly enhancing network throughput.

\end{abstract}

\begin{IEEEkeywords}
Distributed Massive MIMO, Distributed Pilot Assignment, Pilot Contamination
\end{IEEEkeywords}

\section{Introduction}
\IEEEPARstart{T}{he} rapid growth and adoption of wireless communication based data services has prompted the development of technologies that can improve the capacity, reliability, and efficiency of cellular networks. 

By reaping the benefits of mMIMO and Ultra Dense Network (UDN), distributed mMIMO is emerging as a promising technology \cite{ngo2017cell,demir2021foundations}. It abandons the concept of distinct cells and deploys a large number of distributed access points (APs) throughout the coverage area \cite{ngo2017cell}. As there are no traditional cells, the inter-cell interference that plagues cellular systems is eliminated. Further, as the user equipments (UEs) are very close to APs, this may provide high coverage probability in the network, thus improving quality-of service to the users. However, as in traditional mMIMO systems, channel estimation is a critical challenge in distributed mMIMO networks. One commonly employed approach is blind convolution techniques, such as matrix decomposition-based signal detection. However, these methods often have high computational complexity. To mitigate this complexity, works have attempted to exploit the channel hardening property of mMIMO systems \cite{aswathylakshmi2023pilotless}. However, in distributed mMIMO, the channel hardening is not always guaranteed \cite{chen2018channel}, so getting the estimates using matrix decomposition-based channel estimation is not a viable option.

Pilot-based channel estimation presents a simple and low-complexity approach to obtain channel estimates in the distributed mMIMO systems. However, before deploying it in networks an associated issue of pilot contamination needs to be addressed. Pilot contamination arises when the same pilot sequences are used for estimating channels of more than one user, leading to estimation errors and subsequent performance degradation. In practical systems it is almost impossible to assign orthogonal pilots to all users, so the users may reuse the pilots. To minimize the impact of pilot contamination, pilot assignment (PA) should be carried out so that it minimizes the contamination. A lot of work on pilot assignment in distributed mMIMO networks exists, but most of it is centralized in nature \cite{ngo2017cell,bjornson2020scalable,liu2020graph,buzzi2020pilot,chen2020structured,zeng2021pilot,chen2022improving,yu2022topological}.
In \cite{ngo2017cell}, the authors propose two pilot assignment schemes, namely, random pilot assignment and greedy pilot assignment based on throughput improvement. In \cite{bjornson2020scalable}, the authors propose a scalable scheme which performs joint pilot assignment and AP-UE association. In \cite{liu2020graph}, the authors have proposed a graph coloring-based pilot assignment, where the AP-UE association takes place initially and then pilots are assigned using graph coloring. If the pilot assignment fails, the AP-UE associations are updated, and so is the pilot assignment. In \cite{buzzi2020pilot}, the authors propose a scheme based on Hungarian matching algorithm, whereas in \cite{zeng2021pilot}, a weighted graph-based approach is explored. A clustering based scalable PA scheme is proposed in \cite{chen2020structured} where UEs in the same cluster are allocated the same pilot. In \cite{chen2022improving}, the authors have used the interference-aware reward, calculated by treating noise as interference, for the pilot assignment and AP-UE association jointly. Along with random pilot assignment \cite{ngo2017cell}, that can also be implemented in a distributed manner, there are a few distributed schemes \cite{kim2022survey,rahmani2022multi}. In \cite{kim2022survey}, the authors propose a survey propagation based distributed PA that incurs high signaling overhead due to messages passing among the feasible groups. Also, for the proposed scheme the computational cost increases drastically as the number of users-to-pilots ratio increases. In \cite{rahmani2022multi}, a distributed multi-agent reinforcement learning-based scheme is proposed. This, however, demands centralized training, resulting in increased signaling overhead, sensitivity to the training data, and lack of explainability of the model.

To address these challenges, we propose a distributed PA scheme. The novelty of the proposed scheme is that it jointly performs the pilot assignment and AP-UE association. With this, orthogonal pilots are allocated to the topmost UEs of each AP, determined by the largest large-scale fading coefficient. Subsequently, each AP independently associates itself with these topmost UEs with different pilots. The proposed scheme not only has low convergence time, but also improves the overall spectral efficiency (SE), and has much lower signaling overhead than the one in \cite{kim2022survey}. We compare the performance of the proposed scheme with existing schemes through extensive numerical simulations and establish its superiority in terms of scalability and overall system throughput. 

\textit{Organization:} The paper is structured as follows: Section \ref{system_model} presents the system model. Section \ref{resource_assignment} introduces the proposed distributed resource allocation scheme. Section \ref{simulation} evaluates performance of the proposed scheme and demonstrates its effectiveness. Finally, Section \ref{conclusion} concludes the paper by summarizing our work and discussing its future directions. 
\section{System Model}
\label{system_model}
We consider a distributed mMIMO network configuration comprising of \textit{T} UEs and \textit{M} APs. The set of UEs is denoted as $\mathcal{T}$, while the set of APs is denoted as $\mathcal{M}$. Each AP is equipped with \textit{A} antennas, while each UE is equipped with a single antenna. Both the APs and UEs are uniformly distributed over the coverage area. To facilitate the coordination and processing of UE signals, the APs are connected to a Central Processing Unit (CPU) through a front-haul connection. To minimize computation cost and front-haul overload, a scalable architecture is employed \cite{bjornson2020scalable}. This architecture leverages the observation that more than 95\% of the received signal strength is concentrated in a few nearby APs \cite{ngo2017total}. We consider the TDD mode for our operations and assume channel reciprocity.

To model the wireless channel, we consider a standard block fading model with a resource block of length $ L_{c} $. The channel coefficient $\textbf{h}_{t,m} \in \mathbb{C^{\textit{A}}}$ represents the spatially correlated Rayleigh fading channel between the $t^{th}$ UE and the $m^{th}$ AP. It follows a complex Gaussian distribution such that  $\textbf{h}_{t,m}{\sim} \mathcal{N}_{\mathbb{C}}(0,\textbf{R}_{t,m})$, where $\textbf{R}_{t,m} \in \mathbb{C}^{\textit{A}\times \textit{A}}$ represents the spatially correlated matrix that incorporates large-scale fading coefficient (LSFC), denoted by $\beta_{t,m}$, that accounts for various factors such as path-loss and shadow-fading and can be calculated by utilizing periodically broadcasted synchronization signals from the UE $t$ to the AP $m$, \cite{bjornson2020scalable}. We assume that these channel vectors of a particular UE for different APs are independent and this assumption is reasonable as APs are distributed over a large area.

Let $L_p$ denote the number of mutually orthogonal pilots such that the length $L_{p}$ of each resource block is used for pilot training and the remaining for information transmission. As number of orthogonal pilots is smaller than the number of UEs, so pilot reuse comes into play. Let $ \mathcal{T}_{p} $ be the subset of UEs sharing the pilot $p$. 
Following \cite{bjornson2020scalable}, the received signal $\textbf{y}_{p,m} \in \mathbb{C^{\textit{A}}}$ at the $m^{th}$ AP when the pilot $p$ is transmitted by UEs  belonging to $\mathcal{T}_{p}$ is given by :
\begin{align}
\label{eq_1}
 \textbf{y}_{p,m} = \sum_{t\in \mathcal{T}_{p}}\sqrt{L_p\tau_t}\textbf{h}_{t,m} + \textbf{n}_m, 
\end{align}
where $\tau_t $ is the uplink transmit power of the UE $t \in \mathcal{T}_{p}$ and $\textbf{n}_m{\sim}\mathcal{N}_{\mathbb{C}}(0,\sigma^2\mathbf{I}_A) $ denotes the noise at the $m^{th}$ AP. The MMSE estimate of channel $\textbf{h}_{t,m}$ is given by
\begin{align}
\label{eq_2}
\hat{\textbf{h}}_{t,m} = \sqrt{L_p\tau_t}\textbf{R}_{t,m}\boldsymbol{\Psi}^{-1}_{t,m}\textbf{y}_{p,m}, 
\end{align}
where $ \boldsymbol{\Psi}_{t,m} = \sum_{t{\in} \mathcal{T}_{p}}\sqrt{L_p\tau_t}\textbf{R}_{t,m} + \sigma^2\mathbf{I}_A$ is the correlation matrix of $\textbf{y}_{p,m}$. The pilot sharing not only leads to poorer channel estimation, but also affects $\hat{\textbf{h}}_{t,m}$, which becomes more correlated and thus increases the interference among the UEs. \\ 
The association of the UE $t$ and the AP $m$ is denoted by a indicator variable $d_{t,m}$, which equals '$1$'  when the UE $t$ is served by the AP $m$, and '$0$' when the UE $t$ is not so served. The received uplink signal at the AP $m$ is given by  
\begin{align}
\label{eq_7}
 \textbf{y}^{u}_{m} = \sum\limits_{t=1}^{T}\tau_{t}\textbf{h}_{t,m}d_{t,m}x_t + \textbf{n}_m.
\end{align}
The payload signal $x_t \in \mathbb{C}$ is the unit power complex signal sent by the UE $t$ and the power associated with it is denoted by $\tau_{t}$.
The combining vector $\textbf{v}_{t,m} \in \mathbb{C}^{\textit{A}}$ is assigned to the UE $t$ by the AP $m$. The estimate of ${x}_t$ is $\hat{x}_t=\sum_{t{\in}T}d_{t,m}\textbf{v}^{H}_{t,m}\textbf{y}^{u}_{m}$.\\
The uplink SE is \cite{bjornson2020scalable}:
\begin{align}
\label{eq_8}
 \mathsf{SE}^{u}_{t} = (1-\frac{L_p}{L_c})\mathbb{E}\{\log_{2}(1+\mathsf{SINR}^{u}_{t})\},
\end{align}
where expectation is with respect to the channel estimates and $\mathsf{SINR}^{u}_{t}$ represents the uplink instantaneous signal to noise plus interference ratio of the $t^{th}$ UE and is given by
\begin{align}
\label{eq_9}
\mathsf{SINR}^{u}_{t} = \frac{\tau_{t}\vert \sum\limits^{M}_{m=1}d_{t,m}\textbf{v}^{H}_{t,m}\hat{\textbf{h}}_{t,m} \vert^2}{\sum\limits^{T}_{j=1, j \neq t}\tau_{j}\vert \sum\limits^{M}_{m=1}d_{t,m}\textbf{v}^{H}_{t,m}\hat{\textbf{h}}_{t,m} \vert^2 + \textbf{z}_{t}},
\end{align}
where $\textbf{z}_t = \sum^{M}_{m=1}d_{t,m}\textbf{v}^{H}_{t,m}(\sum^{T}_{t=1}\tau_{t}\textbf{C}_t + \sigma^2\mathbf{I}_{MA})\textbf{v}_{t,m}$, and $\textbf{C}_t$ is the error correlation matrix for the collective channel of the UE $t$. 
The precoding vector $\textbf{s}_{t,m} {\in} \mathbb{C}^A$, assigned to the $t^{th}$ UE by the $m^{th}$ AP, is defined as in \cite{bjornson2020scalable}, and the received downlink signal at the $t^{th}$ UE is:
\begin{align}
{y}^{d}_{t} = \sum\limits_{m=1}^{M}\textbf{h}^{H}_{t,m}\sum\limits_{j=1}^{T}d_{j,m}\textbf{s}_{j,m}x_j + n_t, \label{eq_4}
\end{align}
where $n_t{\sim}\mathcal{N}_{\mathbb{C}}(0,\sigma^2) $ is the thermal noise, $x_j \in  \mathbb{C}$ is unit power complex signal sent to the $t^{th}$ UE and $\tau_{t,m}$ is the downlink power assigned to the $t^{th}$ UE by the $m^{th}$ AP, allocated using the fraction power allocation \cite{bjornson2020scalable}.
By utilizing the use-and-then-forget bound \cite{larsson2014massive}, the downlink SE is
\begin{align}
\label{eq_5}
 \mathsf{SE}^{d}_{t} = (1-\frac{L_p}{L_c})\log_{2}(1+\mathsf{SINR}^{d}_{t}),
\end{align}
where $\mathsf{SINR}^{d}_{t}$ represents the downlink instantaneous signal to noise plus interference ratio of the $t^{th}$ UE and is given by
\begin{align}
\label{eq_6}
\mathsf{SINR}^{d}_{t} = \frac{\vert\mathbb{E}\{{\sum\limits^{M}_{m=1}d_{t,m}\textbf{h}^{H}_{t,m}\textbf{s}_{t,m}}\} \vert^2}{\textbf{z} -\vert\mathbb{E}\{{\sum\limits^{M}_{m=1}d_{t,m}\textbf{h}^{H}_{t,m}\textbf{s}_{t,m}}\} \vert^2 + \sigma^2},
\end{align}
where $\textbf{z}= \sum^{T}_{j=1}\vert\mathbb{E}\{{\sum^{M}_{m=1}d_{j,m}\textbf{h}^{H}_{j,m}\textbf{s}_{j,m}}\} \vert^2$ and expectation, as before, is with respect to the channel realizations.

\section{Distributed Pilot Allocation}
\label{resource_assignment}
We propose a distributed pilot assignment scheme to address limited availability of mutually orthogonal pilots ($L_p {<} T$) and mitigate pilot contamination. A key prerequisite for distributed mMIMO is that each UE be served by at least one AP. Additionally, each AP must assign a given pilot to at most one UE to minimize pilot contamination.
To guarantee that these requirement are met, a specific AP is designated as the controller-AP for each UE. Therefore, in the proposed scheme first, a controller-AP is assigned to each UE in a distributed manner. Then, pilots are assigned distributively to each UE by its associated controller-AP. Furthermore, to fully leverage the benefits of mMIMO and make the system scalable, each UE needs to be connected to an adequate number of APs. To achieve this, a distributed AP-UE clustering scheme is employed in the final step based on the constraint that each AP serves one UE per pilot\footnote{The number of users can grow up to a certain extent such that $T < MA$, for the system to behave as mMIMO \cite{ngo2017cell}. Therefore, the above assumption does not hinder the performance when numbers of users is very large.}, thus minimizing pilot contamination while providing significant power to the UEs. 
 
For each AP $m$, we define a subset $\mathcal{U}_m$ of UEs that includes, at most, the leading $L_p$ UEs of the $m^{th}$ AP, in decreasing order of the parameter $\beta_{t,m}$. Let $\mathcal{T}_{m}^{c}$ be the subset of UEs with the AP $m$ as their controller-AP and $\mathcal{L}_{t}^p$ be the set of available pilots for UE $t$, initialized with all $L_p$ pilots.

\subsection{Controller-AP assignment} 
For the controller-AP assignment, we propose a distributed algorithm. In the proposed approach, each UE is responsible for selecting its controller-AP from the set of available APs using their respective LSFC values, by following the standard random access procedure \cite{sesia2011lte} with the broadcast signal \cite{bjornson2020scalable}. However, there is a challenge when more than $L_p$ UEs intend to be allocated a particular AP as their controller-AP. This challenges our design requirement that each AP can serve at most one UE per pilot. Further, if a UE requests an AP with a lower LSFC value to be its controller-AP, this could  potentially lead to degradation in the system performance. To address these, we propose a Controller-AP selection algorithm, where each UE is assigned one AP from the set $\mathcal{M}_{t}^{c}$ of APs as its controller-AP. Also, an AP can serve as the controller-AP of at most $L_p$ UEs. The details of constructing subset $\mathcal{M}_{t}^{c}$ are given in Algorithm \ref{C_AP_selection}. In order to identify the APs that are serving more than $L_p$ UEs and require some of their associated UEs to request a new controller-AP assignment, and the APs that are just serving $L_p$ UEs and so are unable to accept controller-AP requests for more UEs, the APs are categorized into two set. The former are called the oversaturated-APs and are denoted as $\mathcal{M}_D$, and the latter are called the inert-APs and are denoted as $\mathcal{M}_I$. The detailed procedure to select controller-AP distributively from the set of potential controller-APs is given by Algorithm \ref{C_AP_selection}. The Algorithm \ref{C_AP_selection} is carried out at the APs in cooperation with the UEs.

\begin{algorithm}[!]
\caption{Controller-AP selection}\label{C_AP_selection}
\begin{algorithmic}[1]
\STATE \textbf{Input:} $\mathcal{M}_{t}^{c}=\emptyset$, $\beta_{t,m}$, $\mathcal{T}_{m}^{c} = \emptyset$, $\mathcal{M}_D = \emptyset$ and $\mathcal{M}_I = \emptyset$ $\forall$ $t{\in}\mathcal{T}$, $m{\in}\mathcal{M}$ 
\STATE \textbf{Returns:} Controller-AP assignment for each UE.
\FOR{each UE $t \in \mathcal{T}$}
       
    \STATE Calculate $\beta^{d}_{t}$ as the difference between the largest and second largest LSFC values of $\beta_{t,m}$.
 \ENDFOR
    \STATE Calculate $\gamma^{th}= ((\beta^{d}_{t})_{max} + (\beta^{d}_{t})_{min})/2$.  
   \FOR{each UE $t \in \mathcal{T}$}
     \STATE   $\mathcal{M}_{t}^{c} =  \mathcal{M}_{t}^{c} \cup  m_{max}, $\ where$\ \underset{m}{m_{max}=\argmax~}\beta_{m,t} $.          
      \FOR{each AP $m \in \mathcal{M}$}
       \IF{ $\beta_{m_{max},t}- \beta_{m,t} <= \gamma^{th}$}
        \STATE   $\mathcal{M}_{t}^{c} =  \mathcal{M}_{t}^{c} \cup  m$.
       \ENDIF
        \ENDFOR
    \STATE Choose a controller-AP randomly from $\mathcal{M}_{t}^{c}$.
    
    \ENDFOR
   
    \FORALL{ AP $m\in\mathcal{M}$ } 
    \STATE \!\!\!\!\!\! Update $\mathcal{T}_{m}^{c} = \mathcal{T}_{m}^{c} \cup t$, where AP m is a controller-AP of UE $t$.
      \IF{ $\vert \mathcal{T}_{m}^{c} \vert > L_p$ }
      \STATE Update $\mathcal{M}_D = \mathcal{M}_D \cup m$.
      \ELSIF{$\vert \mathcal{T}_{m}^{c} \vert = L_p$}
      \STATE Update $\mathcal{M}_I = \mathcal{M}_I \cup m$.
      \ENDIF
        
    \ENDFOR
    \WHILE{$\mathcal{M}_D \neq \emptyset$}
     \FOR{each AP $m{\in}\mathcal{M}_D$ }
     \STATE Arrange $t \in \mathcal{T}_{m}^{c} $ in increasing order of $|\mathcal{M}_{t}^{c}|$, where priority is given to a UE with a larger LSFC w.r.t AP $m$ to form a new set $\mathcal{T}_{new}$.
     \STATE Set $\mathcal{T}_{m}^{c} = \emptyset$ and include first $L_p$ UEs from $\mathcal{T}_{new}$ in $\mathcal{T}_{m}^{c}$.
     
     \STATE AP $m$ unassigns itself as the controller-APs of all UEs in $(\mathcal{T}_{new}-\mathcal{T}_{m}^{c})$.
     \FOR{ each UE $t{\in}(\mathcal{T}_{new}-\mathcal{T}_{m}^{c})$}
     \STATE Update $\mathcal{M}_{t}^{c} \leftarrow \mathcal{M}_{t}^{c} - \mathcal{M}_I -m$.
     \STATE Choose a controller-AP randomly from $\mathcal{M}_{t}^{c}$.
     \ENDFOR
     \ENDFOR
     \STATE Set $\mathcal{M}_D = \emptyset$, $\mathcal{M}_I = \emptyset$, $\mathcal{T}_{m}^{c} = \emptyset$ and repeat steps in lines 16-23.
      
     \ENDWHILE
\end{algorithmic}
\end{algorithm}

\subsection{Distributed Pilot assignment}
Now we propose a distributed pilot assignment scheme. In order to minimize pilot contamination, we introduce a contamination matrix $\textbf{Ad}$, a binary indicator matrix, indicating whether a pair of UEs can share a pilot or not. Every AP assigns distinct pilots to its serving UEs and informs\footnote{The exchange of information about pilots among APs is carried out via the CPU.} its neighboring APs in $\mathcal{M}_{m}^{nb}$ about its pilot assignment. The construction of $\mathcal{M}_{m}^{nb}$ is explained in Algorithm \ref{PA_assignment}. If a pair of UEs have the same pilot and $\textbf{Ad}$ indicates that these two UEs cannot share the pilot, then the AP allows a UE to retain the pilot if either the cardinality of set $\mathcal{L}^{p}_t$ for this UE is smaller or its index is smaller than that for the other UE. The detailed procedure for assigning pilots to UEs is outlined in Algorithm \ref{PA_assignment}. The Algorithm \ref{PA_assignment} is carried out at the APs.

\begin{algorithm}[!t]
\caption{Distributed Pilot Assignment}\label{PA_assignment}
\begin{algorithmic}[1]
\STATE \textbf{Input:}$\mathcal{M}_{m}^{nb}= \emptyset$, the set of neighboring APs, $\mathcal{T}_{m}^{c}$, $\mathcal{L}^{p}_t$, $\mathcal{U}_m$, $count = 0$, $max\_iter$  and  $a_{i,j}=0, \forall i,j,t{\in}\mathcal{T}$  and $ m \in \mathcal{M}$.
\STATE \textbf{Returns:} Pilot assignment for each UE.
\FOR{each AP $m \in \mathcal{M}$ }
  \FOR{$t \in \mathcal{U}_m$}
   \IF{$t \in \mathcal{U}_m - \mathcal{T}_{m}^{c}$}
    \STATE Update $\mathcal{M}_{m}^{nb} = \mathcal{M}_{m}^{nb} \cup m^{c}_{t}$, where $m^c_t$ represents the controller-AP of UE $t$.
    \STATE Update $\mathcal{M}_{m^{c}_{t}}^{nb} = \mathcal{M}_{m}^{nb} \cup m$.
        \ENDIF
        \ENDFOR
        \STATE For UEs $t,t'{\in}\mathcal{U}_m, where \ t\neq t'$, then $\textbf{Ad}\{t,t'\}=1$. 
 \ENDFOR

    \WHILE{ any UE $t \in \mathcal{T}$ is not assigned a pilot and $\mathcal{L}_{t}^p \neq \emptyset$}
     \FOR{ each AP $m$ }
     \FOR{ each unassigned UE $t \in \mathcal{T}_{m}^{c}$}
     \STATE AP $m$ assigns distinct but unused pilots from $\mathcal{L}_{t}^p$.
        
      \ENDFOR
     \ENDFOR
     \STATE Each AP $ m \in \mathcal{M} $ informs the other APs in $\mathcal{M}_{m}^{nb}$ about the pilots of UEs in $\mathcal{T}_{m}^{c}$. 
     \FOR{each AP $m \in \mathcal{M}$ }
      \FOR{$t \in \mathcal{U}_m$}
        \FOR{$t' \in \mathcal{U}_m$}
        \IF{$t$ and  $t'$ have the same pilot}
           \IF {$|\mathcal{L}_{t}^p| >|\mathcal{L}_{t'}^p|$}
           \STATE The UE $t'$ is allowed to keep its pilot.
           \ELSIF {$|\mathcal{L}_{t}^p| =|\mathcal{L}_{t'}^p|$}
             \STATE The UE with lower index between $t$ and $t'$ is allowed to keep its pilot.
             \ENDIF
             \ENDIF
             \ENDFOR
             \ENDFOR
             
             \ENDFOR

     \STATE Each AP $ m \in \mathcal{M} $ again informs the other APs in $\mathcal{M}_{m}^{nb}$ about the pilots of UEs in $\mathcal{T}_{m}^{c}$.
     \FOR{ each AP $m$ }
     \FOR{each UE $t \in \mathcal{T}_{m}^{c}$}
     
     \STATE Updates the $\mathcal{L}_{t}^p$ by removing the pilot of UE $t'$ for which $\mathbf{Ad}\{t',t\}=1, \ \forall t'\in \mathcal{T}$.
     \ENDFOR
     \ENDFOR 
     
     \ENDWHILE
     \IF{ any UE $t$ left unassigned}
      \STATE The AP $m^{c}_{t}$ assigns an unused pilot to the UE $t$.
      \ENDIF

\end{algorithmic}
\end{algorithm}
%


\subsection{AP-UE Clustering}
The distributed UE-AP clustering is performed to improve the spectral diversity, as in \cite{bjornson2020scalable}. Each AP serves at most one UE per pilot in order to minimize the pilot contamination. Initially, each AP $m$ chooses to serve each UE $t \in \mathcal{T}_{m}^{c} $. Then the AP $m$, for each pilot (except pilots of UEs in $ \mathcal{T}_{m}^{c} $), chooses a UE to serve based on the largest LSFC. 

\subsection{Complexity analysis}
The complexity of graph-coloring based pilot assignment algorithm in \cite{liu2020graph} is $\mathcal{O}(\textit{T}\textit{M}log_2\textit{M}+2\textit{T}\textit{M}+\textit{T}^2)$. The complexity of scalable PA in \cite{bjornson2020scalable} is $\mathcal{O}(\textit{T}\textit{M}+\textit{T}\textit{L}_p + \textit{M}\textit{L}_p)$. The complexity of IAR-based PA in  \cite{chen2022improving} is $\mathcal{O}(\textit{T}^2\textit{M} +\textit{M}\textit{L}^{3}_p + \textit{T}\textit{M}^2)$.

The complexity of our proposed Algorithm \ref{C_AP_selection} mainly depends upon the execution in lines 24-35, the \textbf{while} loop iterates at most \textit{M} times, given that any AP can be a unsaturated AP at most once, and also the \textbf{for} loop in line 25-33 can be executed in parallel with each AP as a separate processor, therefore its worst-case complexity is $\mathcal{O}(\textit{T}\textit{M})$. The complexity of Algorithm \ref{PA_assignment}, where each AP acts as a separate processor, depends on finding the set $\mathcal{M}_{m}^{nb}$ for each AP. This has complexity of $\mathcal{O}(\textit{L}^{2}_p)$. The Algorithm \ref{PA_assignment} also depends on execution of the \textbf{while} loop between lines 12-38. Assuming the \textbf{while} loop runs $I$ times, the complexity of execution of the \textbf{while} loop is $\mathcal{O}(\textit{I}\textit{L}^{2}_p)$. Therefore, the overall complexity of Algorithm \ref{PA_assignment} is $\mathcal{O}(\textit{L}^{2}_p+\textit{I}\textit{L}^{2}_p)$. Putting everything together, the overall complexity of the proposed scheme including AP-UE association is $\mathcal{O}(\textit{T}\textit{M}+\textit{L}^{2}_p +\textit{I}\textit{L}^{2}_p+\textit{L}_p)$ or $\mathcal{O}(\textit{T}\textit{M}+\textit{I}\textit{L}^{2}_p)$.

\subsection{Overhead analysis}
During the controller-AP selection there is communication between the APs and UEs, i.e., the UEs request APs to be their controller-AP and each AP respond to each UE whether it can serves as its controller-AP or not. This adds to the overall communication overhead. Also, during the pilot assignment phase the APs inform each other via the CPU about the pilots they are using to serve their respective UEs.

\section{Performance Evaluation}
\label{simulation}
For numerical simulations we have considered an area of $2000\times 2000$ square meters in which \textit{T} UEs  and 100 APs are uniformly distributed. Each AP is equipped with 4 antennas. We have considered APs to be 10 meters higher than the UEs. We set $L_c = 200,\  L_p = 10, \ \tau_t=100 \ mW $ and $1000 \ mW$ downlink power of each AP. We have considered bandwidth to be 20 MHz. To simulate the large scale propagation model, we have considered the 3GPP Urban Microcell scenario and  the local scattering model for spatial correlation same as in \cite{bjornson2017massive}.  We have averaged our results over 50 network instances and for each instance we have considered 500 channel realizations. For all the simulations, P-MMSE precoder \cite{bjornson2020scalable} is used. For spectral efficiency comparison we have considered centralized schemes, such as scalable PA \cite{bjornson2020scalable}, IAR-sum PA \cite{chen2022improving}, graph-coloring-based PA \cite{liu2020graph}; and a distributed scheme: survey-propagation-based (survey) PA \cite{kim2022survey}. For a fair comparison, in scalable and survey-propagation, a AP can serve at most $L_p$ UEs.

\subsection{Downlink Operations} 
\subsubsection{Comparison with centralized schemes}
\begin{figure}[!h]
\centering
\includegraphics[width=\textwidth]{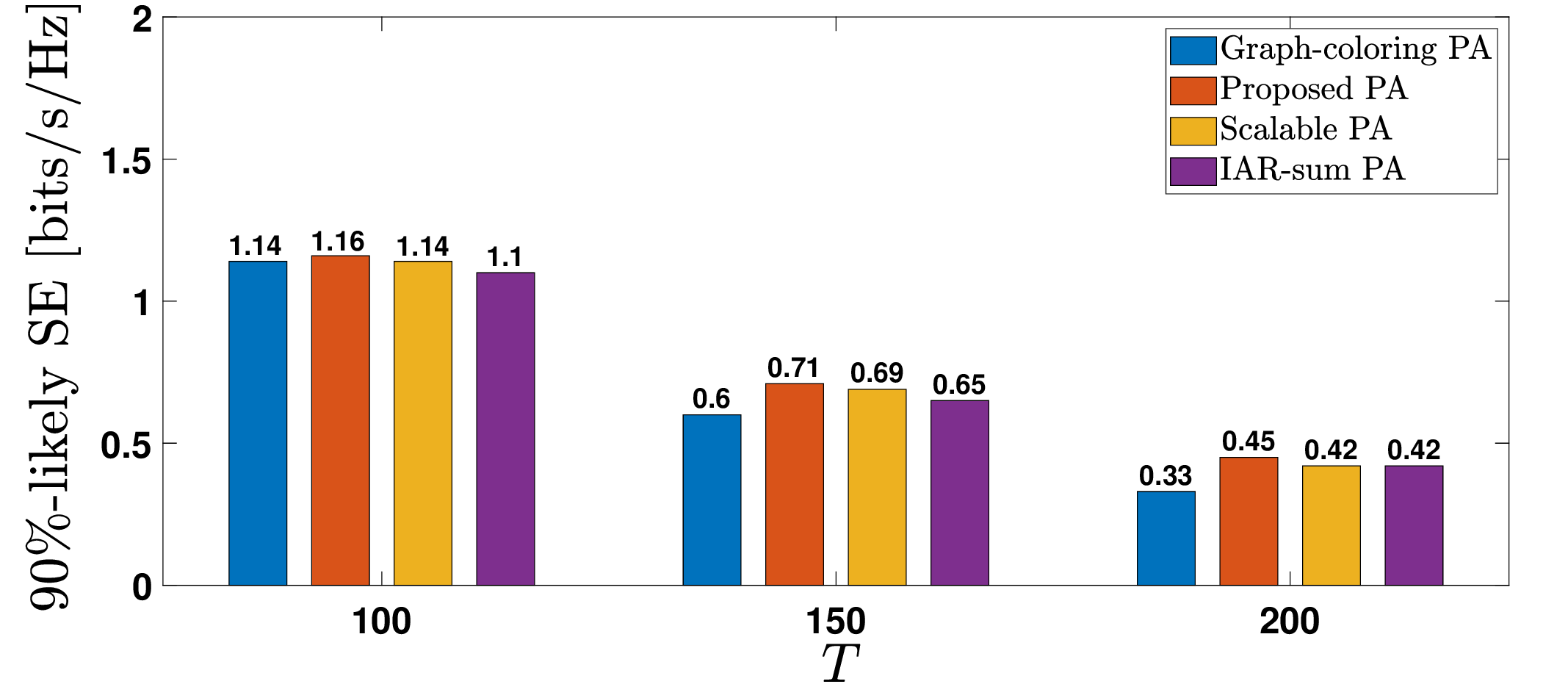}
\caption{90\%-likely downlink SE vs $T$.}
\label{fig_7}
\vspace{-0.1in}
\end{figure}

Fig.~\ref{fig_7} shows the plot of the 90\%-likely downlink SE for different numbers of UEs.  The proposed distributed scheme outperforms the scalable PA scheme by 2\%, the graph-coloring PA scheme by 2\%, and the IAR-sum PA scheme by 6\% in scenario involving 100 UEs. Similarly, improvements with 150 UEs are 3\%, 18\%, and 9\% respectively, and improvements with 200 UEs are 7\%, 36\%, and 7\%, respectively. It can also be observed that as the number of UEs increases, the 90\%-likely downlink SE of all schemes decreases, but our proposed scheme sees a lesser decrease than all other schemes. These results demonstrates the superiority of our proposed scheme in terms of 90\%-likely downlink SE than all other schemes for all user scenarios. 

\begin{figure}[h]
\centering
\includegraphics[width=\textwidth]{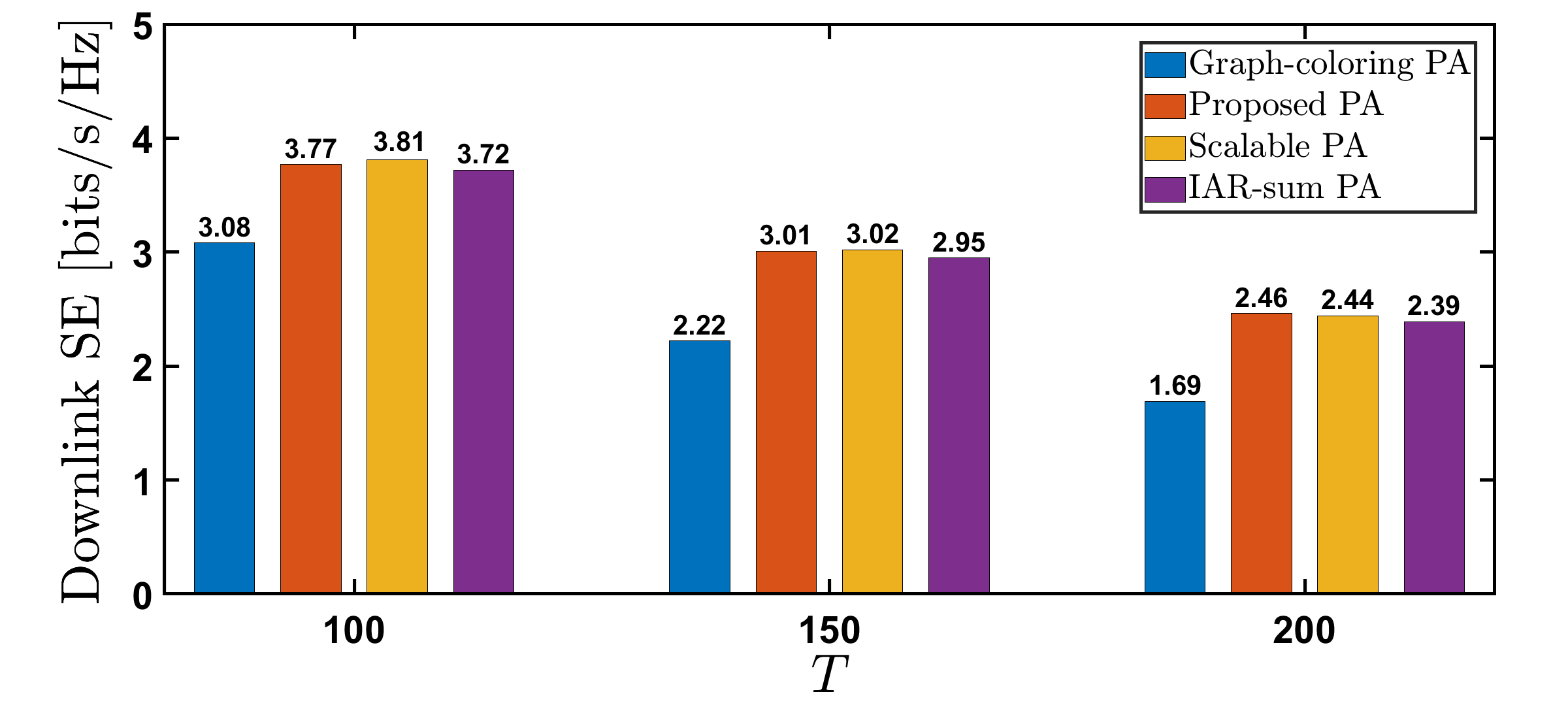}
\caption{Average downlink SE vs $T$.}
\label{fig_5}
\vspace{-0.1in}
\end{figure}

Similarly Fig.~\ref{fig_5} shows the graph between average downlink SE for different numbers of UEs. For $T=100$, our proposed scheme demonstrates superior performance compared to the graph-coloring PA and IAR-sum PA by 22\% and 1\% respectively. However, it is slightly outperformed (by 1\%) by the scalable PA. For $T=150$, the proposed scheme outperforms both the graph-coloring based PA and the IAR-sum PA, while achieving comparable performance to the scalable PA. Finally, for $T=200$, the proposed scheme surpasses all other schemes.

\subsubsection{Comparison with a distributed scheme}
\begin{figure}[!h]
\centering
\includegraphics[width=\textwidth]{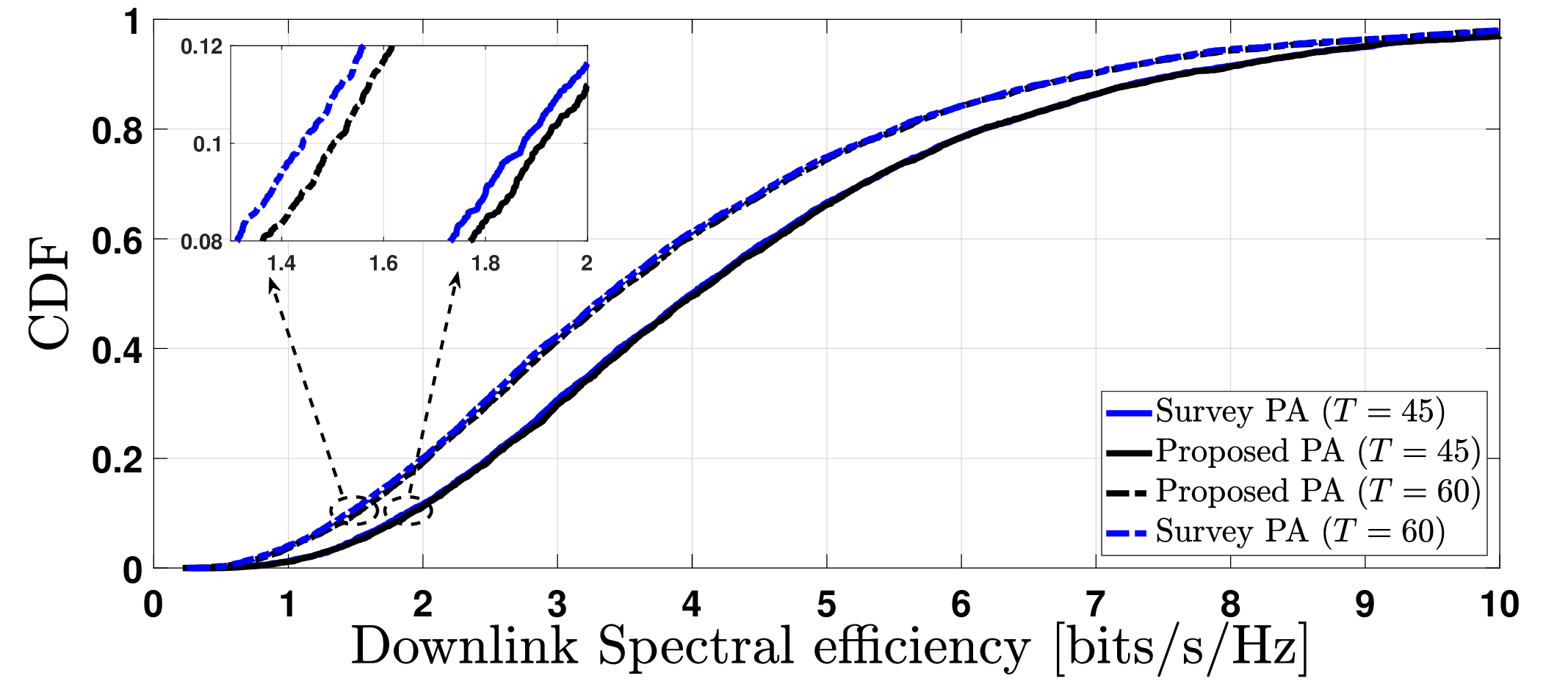}
\caption{CDF of downlink SE, when $\frac{T}{L_p}=3$.}
\label{fig_3}
\vspace{-0.1in}
\end{figure}

Fig.~\ref{fig_3} shows the CDF of 90\%-likely downlink SE for distributed PA schemes. To decrease the computational cost of the distributed survey PA scheme, we have considered the number of UEs to the number of pilots ratio to be $3$. The plot shows that the proposed scheme outperforms the survey-propagation-based scheme, thus demonstrating its superiority.

\subsection{Uplink Operations}
\subsubsection{Comparison with centralized schemes}
\begin{figure}[!h]
\centering
\includegraphics[width=\textwidth]{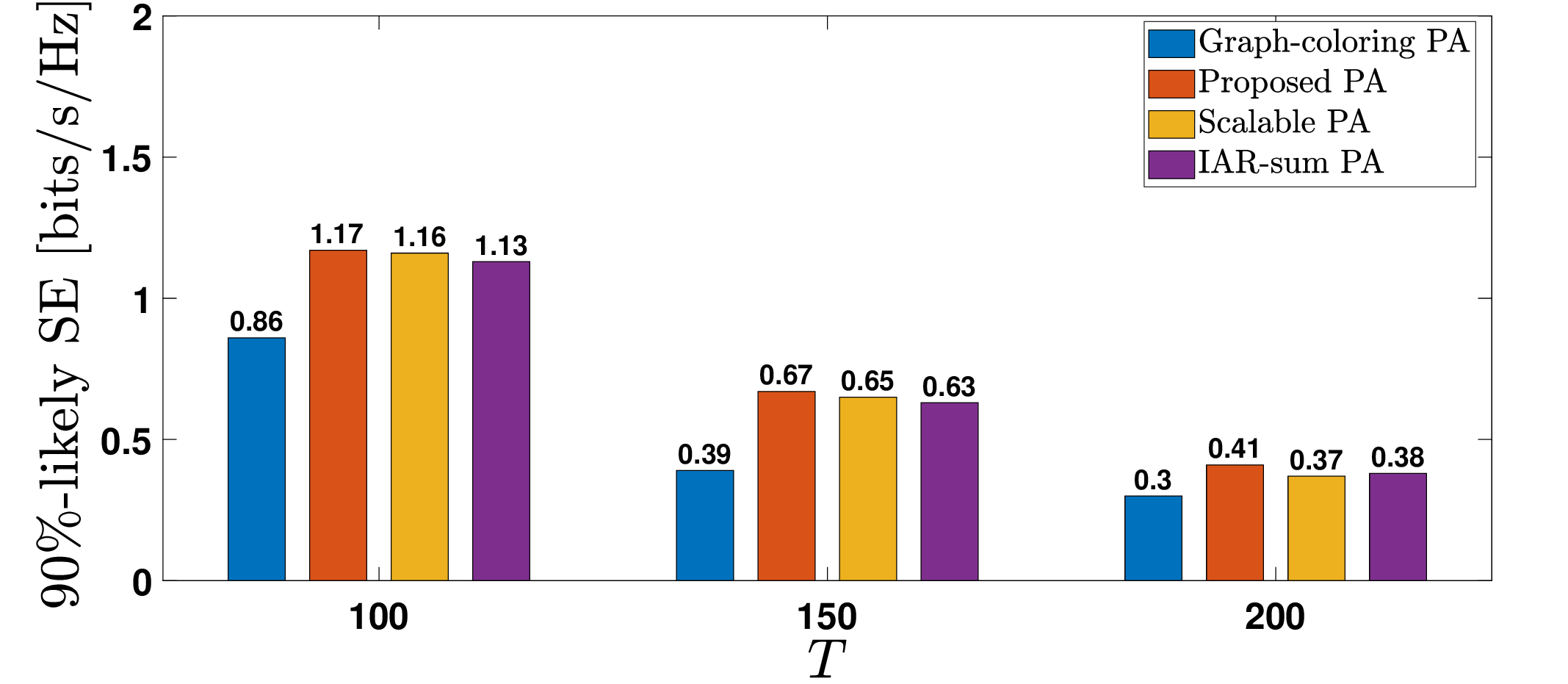}
\caption{90\%-likely uplink SE vs $T$.}
\label{fig_8}
\vspace{-0.1in}
\end{figure}

Fig.~\ref{fig_8} shows the plot of the 90\%-likely uplink spectral efficiency for different numbers of UEs. In a scenario involving 100 UEs, the proposed approach outperforms the scalable PA technique by 1\%, the graph-coloring-based PA scheme by 36\%, and the IAR-sum PA scheme by 4\%. Similarly, with 150 UEs, improvements are 3\%, 72\%, and 6\%, respectively, and improvements with 200 UEs are 11\%, 105\%, and 9\%, respectively. It can also be seen that when the number of UE increases, the 90\%-likely downlink SE of all schemes decreases, however our proposed scheme has a lesser decrease than all other schemes.

\begin{figure}[!h]
\centering
\includegraphics[width=\textwidth]{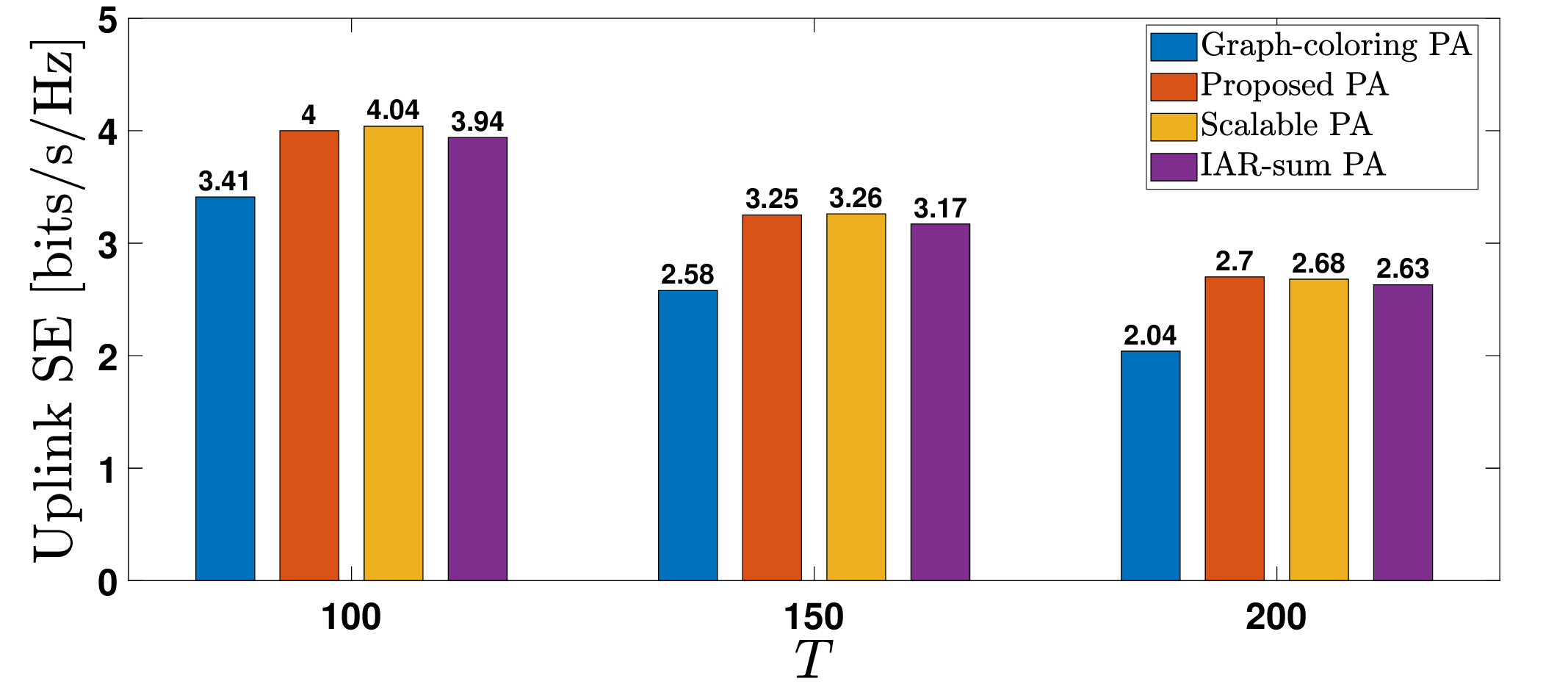}
\caption{Average uplink SE vs $T$.}
\label{fig_6}
\vspace{-0.1in}
\end{figure}

Fig.~\ref{fig_6} shows the graph between average uplink SE for different numbers of UEs. For $T=100$, our proposed scheme demonstrates superior performance compared to the graph-coloring PA by approximately 17\% and the IAR-sum PA by nearly 2\%. It is, however, outperformed (by $\le$ 1\%) by the scalable PA. The proposed scheme surpasses both the graph-coloring-based PA and the IAR-sum PA by a significant margin for $T=150$, while matching the scalable PA performance.  Finally, for $T=200$, our proposed scheme surpasses all other schemes.

\subsubsection{Comparison with a distributed scheme}
\begin{figure}[!h]
\centering
\includegraphics[width=\textwidth]{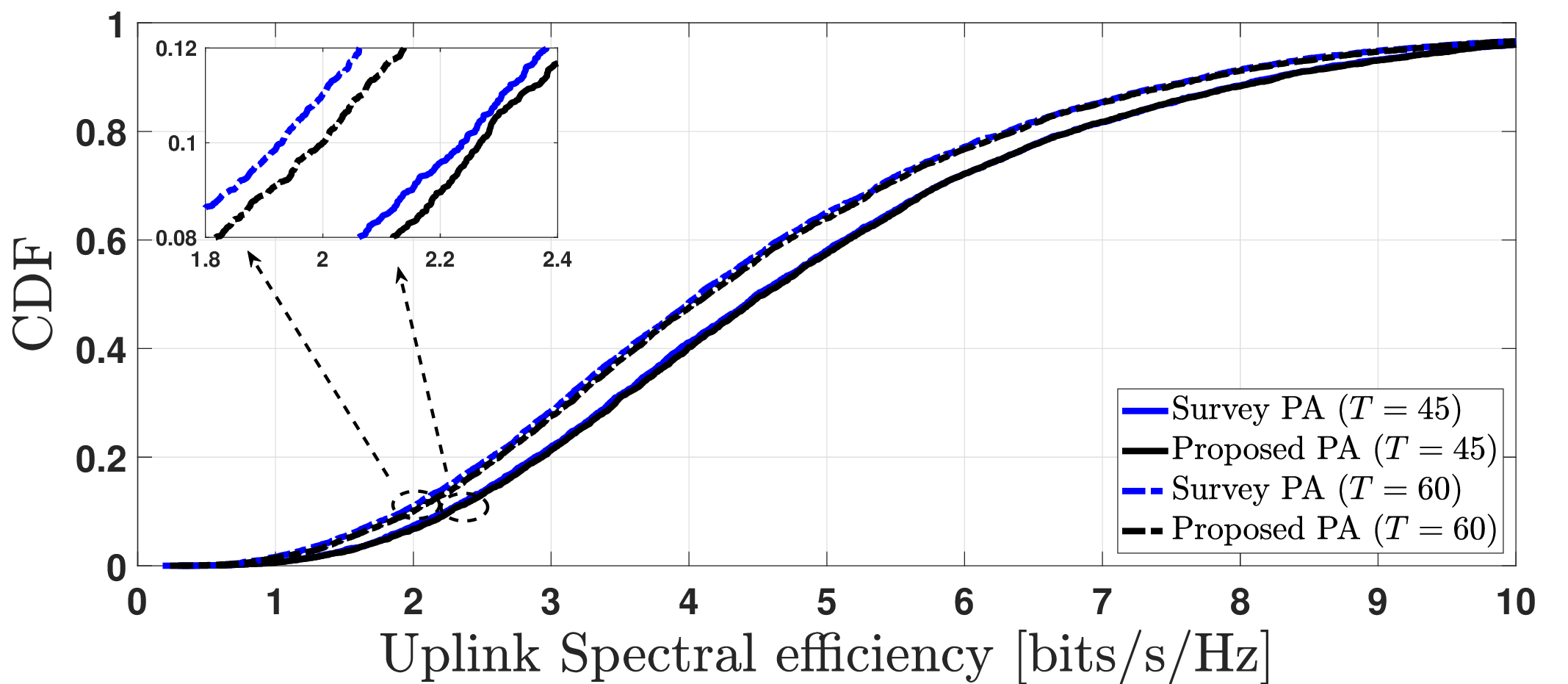}
\caption{CDF of uplink SE, when $\frac{T}{L_p}=3$.}
\label{fig_4}
\vspace{-0.1in}
\end{figure}

Fig.~\ref{fig_4} shows the uplink SE plot for distributed PA schemes. Again in order to decrease the computational costs in distributed survey PA scheme, we have considered the number of UEs to the number of pilots ratio to be $3$. In terms of 90\%-likely SE, the proposed scheme outperforms the survey-propagation-based scheme by 2\% for $T=45$ and by nearly 4\% for $T=60$. Thus, indicating the superiority of our proposed scheme.

\subsection{Discussion}
Superior performance of our proposed distributed scheme in terms of both 90\%-likely downlink as well as uplink SE demonstrates its ability to outperform not only distributed survey PA scheme, but all of the aforementioned centralized schemes, and its performance continues to improve as the number of UE grows. This is primarily attributable to our approach's priority of reducing pilot contamination for each UE by assigning distinct pilots to UEs that are more susceptible to contamination, thus improving fairness among UEs in terms of contamination. Although our proposed scheme does not focus on improving the average SE, it still achieves greater average downlink as well as uplink SE than IAR-sum and graph-coloring PA schemes in all scenarios. However, our approach has a modest disadvantage in terms of both average downlink and uplink SEs over the scalable PA scheme for low user density scenarios. This minor difference in average SE demonstrates that, despite being distributed, our approach is competitive with centralized approaches.

\newpage

\section{Conclusion}
\label{conclusion}
The proposed distributed pilot assignment technique in distributed mMIMO systems may substantially enhance spectral efficiency compared to centralized as well as existing distributed schemes. By using a distributed approach with parallel execution, this strategy not only boosts spectral efficiency but also reduces convergence time and increases the network's fault-tolerance as well as scalability. Our results highlight the importance of using distributed resource allocation algorithms to achieve optimal performance in distributed mMIMO systems. Further, those illustrate the potential benefits of adopting distributed approaches in future distributed mMIMO network plans and deployments.


\end{document}